**RESEARCH**

**Open Access**

# Low-intensity pulsed ultrasound promotes mesenchymal stem cell transplantation-based articular cartilage regeneration via inhibiting the TNF signaling pathway

Yiming Chen[1†], Huiyi Yang[2†], Zhaojie Wang[2,4†], Rongrong Zhu[2,4], Liming Cheng[2*] and Qian Cheng[1,2,3*]

## Abstract

**Background**  Mesenchymal stem cell (MSC) transplantation therapy is highly investigated for the regenerative repair of cartilage defects. Low-intensity pulsed ultrasound (LIPUS) has the potential to promote chondrogenic differentiation of MSCs. However, its underlying mechanism remains unclear. Here, we investigated the promoting effects and mechanisms underlying LIPUS stimulation on the chondrogenic differentiation of human umbilical cord mesenchymal stem cells (hUC-MSCs) and further evaluated its regenerative application value in articular cartilage defects in rats.

**Methods**  LIPUS was applied to stimulate cultured hUC-MSCs and C28/I2 cells in vitro. Immunofluorescence staining, qPCR analysis, and transcriptome sequencing were used to detect mature cartilage-related markers of gene and protein expression for a comprehensive evaluation of differentiation. Injured articular cartilage rat models were established for further hUC-MSC transplantation and LIPUS stimulation in vivo. Histopathology and H&E staining were used to evaluate the repair effects of the injured articular cartilage with LIPUS stimulation.

**Results**  The results showed that LIPUS stimulation with specific parameters effectively promoted the expression of mature cartilage-related genes and proteins, inhibited TNF-α gene expression in hUC-MSCs, and exhibited anti-inflammation in C28/I2 cells. In addition, the articular cartilage defects of rats were significantly repaired after hUC-MSC transplantation and LIPUS stimulation.

**Conclusions**  Taken together, LIPUS stimulation could realize articular cartilage regeneration based on hUC-MSC transplantation due to the inhibition of the TNF signaling pathway, which is of clinical value for the relief of osteoarthritis.

†Yiming Chen, Huiyi Yang, Zhaojie Wang have contributed equally to this work and share first authorship.

*Correspondence:
Liming Cheng
limingcheng@tongji.edu.cn
Qian Cheng
q.cheng@tongji.edu.cn
Full list of author information is available at the end of the article

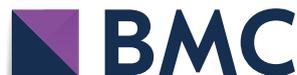

## Background

Articular cartilage, providing a smooth surface and decreasing friction for articulation, suffers great pressure during the frequent daily joint movement and tends to gradually wear away or suddenly get injured [1]. This problem is exacerbated by the poor intrinsic healing and repair capacity of the cartilage, which results in an inability to heal on its own once damaged, thus commonly leading to osteoarthritis (OA) or even lifelong disability without effective treatment [2]. Various cartilage repair procedures have been reported, among which microfracture surgery and autologous osteochondral transplantation are invasive and have limited repair promotion capabilities [3–6]. Autologous chondrocyte implantation (ACI) is a promising therapeutic strategy for cartilage defects that reduce immunologic rejection and minimize the likelihood of transmitting infectious diseases; however, it is limited by an insufficient number of chondrocytes [7–9]. Meanwhile, stem cell transplantation provides a more readily accessible source of cells for the treatment and maintaining multipotency during culture expansion [10], thus overcoming the lack of chondrocytes, and is a promising therapy for cartilage defect repair [11, 12]. Specifically, mesenchymal stem cells (MSCs) have the advantages of rich sources, strong proliferative and self-renewal ability, and multi-lineage differentiation ability including chondrogenic differentiation [13–15], making them an ideal cell type for cell transplantation-based cartilage tissue repair. Studies have shown that MSCs are feasible and safe for use in the repair of cartilage defects [14, 16–19]. Notably, human umbilical cord mesenchymal stem cells (hUC-MSCs) have shown stronger proliferation and differentiation ability compared with human bone mesenchymal stem cells (hBMSCs) and human adipose-derived mesenchymal stem cells (hAD-MSCs) [20].

Ultrasound, as a non-invasive and affordable technology, is widely used in both diagnostic and therapeutic medicine. Through thermal and non-thermal mechanisms, ultrasound can produce various biological effects in vitro and in vivo [21], including mechanical effect, tissue temperature rise, acoustic cavitation, and acoustic streaming. Low-intensity pulsed ultrasound (LIPUS) is a form of ultrasound delivered at an intensity lower than 3 W/cm$^2$ and transmitted through and into tissues as pulsed acoustic pressure waves; it removes the thermal component at higher intensities and primarily provides mechanical stimulation. As an adjuvant physical therapy, LIPUS has been demonstrated to promote the differentiation of stem cells [22, 23], inhibit inflammatory responses [24], accelerate soft tissue regeneration [25, 26], modulate neuronal activity [27, 28], and improve bone healing [29, 30]. In particular, LIPUS has been extensively exploited in tissue engineering research, including cartilage regeneration [31–35].

Furthermore, LIPUS stimulation has been reported to promote the proliferation and self-renewal capacity of MSCs [36], facilitate the differentiation of MSCs into chondrocytes [31, 37–40], and promote the proliferation and matrix production of chondrocytes [41, 42]—suggesting a potential to promote MSC-based cartilage regeneration. Regarding the underlying mechanisms of these bioeffects, previous studies have suggested various possibilities, including upregulation of the expression of cyclin-D1 [40] and regulation of autophagy [38, 43], the MAPK/ERK pathway [44–46], the integrin/PI3K/Akt pathway [41], and the calcium pathway [46].

These findings strongly indicate the potential of LIPUS in regenerating damaged cartilage via MSCs. However, the mechanism underlying these biological effects along with the chondrogenic proliferation and differentiation by LIPUS stimulation has not been fully understood, which limits its further clinical application in repairing injured articular cartilage.

In this study, the effects of LIPUS stimulation on the chondrogenic differentiation of hUC-MSCs, as well as the proliferation and anti-inflammation of human chondrocyte C28/I2 cells, were explored. LIPUS stimulation with various sound intensities ranging from 30 to 250 mW/cm$^2$ was applied. Flow cytometric analysis, Alcian blue staining, immunofluorescence staining, and qPCR analysis were used to detect the cell cycle phases and apoptosis of hUC-MSCs and C28/I2 cells, mature chondrocyte-related marker genes or protein expression, and inflammation-related gene expression. In addition, transcriptome sequencing was used to clarify the mechanism of the key signaling pathway of LIPUS stimulation-induced hUC-MSC chondrogenic differentiation. Moreover, rat knee articular cartilage defect models were established to verify LIPUS- and hUC-MSC-based cartilage regeneration in vivo. Bright-field photography and histological staining with hematoxylin–eosin (H&E) and Safranin O-fast green were used to identify the healing effect on defective cartilage. Although the concept of "LIPUS-stimulated and hUC-MSC-mediated cartilage regeneration" has been previously reported, the present study validated in vivo experiments in injured articular cartilage rat models and proposed a new possible regenerative mechanism related to LIPUS-induced TNF signaling pathway inhibition. Therefore, our present study demonstrates solvable therapeutics of ultrasound-promoted articular cartilage repair based on hUC-MSC transplantation.

## Materials and methods
### LIPUS stimulation
The LIPUS stimulation setup includes a self-developed function generator that simultaneously supports six-channel output and 1 MHz ultrasound transducers with a diameter of 30 mm. The LIPUS used in the experiments was characterized as follows: frequency = 1 MHz, pulse duration = 600 ms, pulse repetition period = 3,000 ms, pulse repetition frequency = 0.33 Hz, duty cycle = 20%, and the spatial-peak temporal-average intensity ($I_{SPTA}$) ranged from 30 to 250 mW/cm$^2$. The acoustic pressure was measured using a calibrated needle hydrophone (ONDA, Sunnyvale, CA, USA), and the effective acoustic pressure in the experiments ranged from 0.05 to 0.14 MPa.

For in vitro cell stimulation experiments, hUC-MSCs or C28/I2 cells were pre-cultured in six-well plates at a density of $2.5 \times 10^4$ cells/mL. Six transducers were placed at the bottom of each well of the six-well plate to generate a 20-min LIPUS stimulation per day. The sham-treated control group (Ctrl group) was also processed as the LIPUS group but without LIPUS stimulation output.

For in vivo rat stimulation experiments, the rats in the LIPUS group were anesthetized with isoflurane, and the knee joint with shaved surface hair was placed on the transducer to receive a 20-min LIPUS stimulation per day. Ctrl group was also anesthetized and placed on the transducer as the LIPUS group but without LIPUS stimulation output.

### Culture of hUC-MSCs
hUC-MSCs used in this experiment were purchased from Cyagen Biotechnology (Guangzhou, China). The cells were cultured with Dulbecco's modified Eagle medium (DMEM)/F12 containing 5% fetal bovine serum (FBS) and 1% penicillin–streptomycin (P/S), placed in a cell incubator at a constant temperature of 37 °C, and continuously maintained at a concentration of 5% $CO_2$. All hUC-MSCs used in the experiments were P2–P4 generations. The culture medium for hUC-MSC chondrogenic differentiation was purchased from Cyagen Biotechnology.

### Culture of C28/I2
Human chondrocytes (C28/I2) used in this experiment were purchased from Procell Life Science & Technology (Wuhan, China). The cells were cultured in DMEM containing 5% FBS and 1% P/S, placed in a cell incubator at a constant temperature of 37 °C, and continuously maintained at a concentration of 5% $CO_2$. The C28/I2 of the generation P0–P2 was used in this study.

### Flow cytometric analysis of hUC-MSC characteristics
To detect the surface markers of hUC-MSCs, the cells were seeded in a six-well culture dish at a density of $2.5 \times 10^4$ cells/mL and cultured for 24 h, and the detect kit was purchased from Cyagen Biotechnology (Guangzhou, China). The cells were digested with trypsin, centrifuged at $300 \times g$ for 5 min, and resuspended in 100 μL PBS, and 2 μL primary antibodies (CD29, CD44, CD73, CD105, CD166, CD11b, CD14, CD34, CD45, and HLA-DR) were added to each tube and incubated for 30 min at 4 °C in the dark. After centrifugation, the cells were washed twice with PBS and incubated with the corresponding secondary antibody at 4 °C for 30 min without light according to the manufacturer's instructions. In this experiment, the fluorescein isothiocyanate (FITC) goat secondary antibody was labeled with CD29, CD44, CD73, CD105, and CD166, and the phycoerythrin (PE) goat secondary antibody was labeled with CD11b, CD14, CD34, CD45, and HLA-DR. Finally, the samples were resuspended in 500 μL flow cytometer buffer and immediately analyzed using a flow cytometer (BD Bioscience, Franklin Lakes, NJ, USA).

### Flow cytometry analysis of apoptosis
To detect the apoptosis of hUC-MSCs and C28/I2 cells after LIPUS treatment, the cells were seeded in a six-well culture dish at a density of $2.5 \times 10^4$ cells/mL and cultured for 24–48 h. The cells were stimulated with LIPUS (at an $I_{SPTA}$ of 0, 30, 50, 70, 100, 150, 200, and 250 mW/cm$^2$ in hUC-MSCs, an $I_{SPTA}$ of 0, 30, 50, 70, 100, and 150 mW/cm$^2$ in C28/I2 cells) for 20 min before the cells were digested after another 2 h of culturing. After dissociating into single cells using trypsin and washing twice with PBS, the cells were fixed in binding buffer (500 μL) containing Annexin V-allophycocyanin (APC) (5 μL) and 7-aminoactinomycin D (7-AAD) (5 μL) for 10 min, and cell apoptosis was detected using a flow cytometer.

### Flow cytometry analysis of cell cycle
To detect the cell cycle phases of hUC-MSCs and C28/I2 cells after LIPUS treatment, the cells were seeded in a six-well culture dish at a density of $2.5 \times 10^4$ cells/mL and cultured for 24 h. The cells were then treated with LIPUS stimulation (at an $I_{SPTA}$ of 0, 30, 50, 70, 100, 150, 200, and 250 mW/cm$^2$ in hUC-MSCs, an $I_{SPTA}$ of 0, 30, 50, 70, 100, and 150 mW/cm$^2$ in C28/I2 cells) for 20 min per day. After 2 days, the treated cells were dissociated into single cells and placed in 70% ice-cold ethanol at 4 °C overnight. The cells were then centrifuged ($300 \times g$, 5 min) and resuspended in a mixture of 100 μL RNase A and 400 μL propidium iodide solution for 30 min, and the cell cycle phase was detected using a flow cytometer.

### Alcian blue staining
The hUC-MSCs were seeded in a 24-well plate at a cell density of $2.5 \times 10^4$ cells/mL and cultured for 24 h to achieve a cell fusion density of 70–80%. After treating the cells with 20 min/day LIPUS stimulation (at an $I_{SPTA}$ of 0, 30, 50, 70, 100, 150, 200, 250 mW/cm$^2$, respectively) for 7 days, the samples were fixed with 4% paraformaldehyde (PFA) for 30 min, aspirated, washed twice with $1 \times$ PBS, and incubated with Alcian blue staining solution for 30 min. The staining solution was aspirated, and the samples were washed twice with $1 \times$ PBS before being transferred to the culture plate to observe the effect of cartilage staining (by cell morphology and staining depth) under an Olympus microscope.

### Establishment of C28/I2 inflammation model
C28/I2 cells were seeded in a six-well plate at a cell density of $5 \times 10^4$ cells/mL. After culturing for 24 h to reach a cell fusion density of 70–80%, lipopolysaccharide (LPS) solution was added to each well at a concentration of 5 μg/mL to establish the inflammatory model, and the stimulation time was 4 h according to our initial assessment. After stimulation, the LPS solution was replaced with fresh C28/I2 culture medium for further analysis.

### Total mRNA extraction, cDNA transcription, and qPCR analysis
To detect the expression of chondrogenic and inflammation-related genes in cells, TRIzol (Takara, Japan) was used to extract total mRNA (500 ng) from the hUC-MSCs and C28/I2 cells, reverse-transcribed into cDNA using PrimeScript RT Kit (Takara), and qPCR was performed in 10 μL reaction volumes (5 μL TB Green, 0.2 μL ROX Reference Dye, 0.2 μL forward/reverse primers, 1 μL cDNA, 3.6 μL H$_2$O) using TB Green Premix Ex Taq (Takara). All qPCR primers used are listed in Additional file 1: Tables S1 and S2.

### Cell cytotoxicity detection
For cell toxicity detection, hUC-MSCs and C28/I2 cells were seeded in a six-well plate at a cell density of $1 \times 10^5$ cells/well, cultured for 24 h to reach a cell fusion density of 70–80%, and treated with 20 min of LIPUS stimulation (at an $I_{SPTA}$ of 0, 30, 50, 70, 100, 150, 200, 250 mW/cm$^2$, respectively) for 7 days. Thereafter, the cells were collected by trypsinization and centrifugation ($300 \times g$, 5 min) and washed thoroughly with assay buffer 2–3 times, before adding 400 μL assay buffer to each sample, followed by the addition of 200 μL staining working solution for 15 min with incubation at 37 °C. The live and dead viability analysis was conducted using a Calcein AM-PI kit purchased from KeyGen. Using an Olympus fluorescence microscope, the live cells (green fluorescence) were detected at 488 nm and the dead cells (red fluorescence) were detected at 594 nm.

### Western blot
For immunostaining of chondrogenic-related proteins, a Western blot analysis was conducted. The hUC-MSCs were counted as $1 \times 10^5$ cells/mL in six-well plates and stimulated with different parameters of LIPUS for 7 days. The cells were collected, and total protein was collected by RIPA lysis; the protein count was determined by a BCA protein quantification kit (KeyGen, Nanjing). The proteins were subject to SDS–PAGE at the same content and transferred into a PVDF membrane. Bovine serum albumin (BSA, 5%) was used to block non-specific binding for 1 h incubation, and the membrane was incubated with primary antibodies of collagen-II, ACAN, SOX-9, and β-actin at 4 °C overnight. After washing thrice with TBST, the membranes were incubated with responsive secondary donkey anti-rabbit of mouse antibodies (1:3000). Finally, the membranes were washed to remove redundant antibodies and exposed using a chemiluminescence imaging system. The specific primary antibodies were listed as follows: collagen-II (Abcam, 1:1,000), ACAN (CST, 1:1,000), SOX-9 (Abcam, 1:1,000), and β-actin (Abcam, 1:3,000).

### Immunofluorescence staining
hUC-MSCs were seeded in a 24-well plate at a cell density of $5 \times 10^4$ cells/mL and cultured for 24 h to achieve a cell fusion density of 70%. Then, the basic culture medium was changed into cartilage differentiation culture medium and the cells were treated for 7 days with 20 min/day LIPUS stimulation (at an $I_{SPTA}$ of 0, 30, 50, 70, 100, and 150 mW/cm$^2$, respectively). The cells were fixed with 4% PFA for 30 min, permeabilized for 10 min, and then incubated with primary antibody overnight at 4 °C, corresponding secondary antibody for 45 min the next day, and with either actin green (1:100) for 20 min, and 4′,6-diamidino-2-phenylindole (DAPI) (1:3,000) for 15 min. The SOX-9 and ACAN antibodies used in this study were purchased from Abcam (Cambridge, UK). Immunofluorescence was observed using a Zeiss confocal microscope (Oberkochen, Germany).

For the mechanistic study, hUC-MSCs were seeded in a 24-well plate at a cell density of $5 \times 10^4$ cells/mL and cultured for 24 h to achieve a cell fusion density of 70%. The cells were divided into Ctrl, TNF-α, LIPUS, and TNF-α & LIPUS groups. The cells in TNF-α and TNF-α & LIPUS groups were pretreated with 10 μg/mL TNF-α for 24 h and then replaced with fresh chondrogenic differentiation medium. LIPUS and TNF-α & LIPUS groups were treated with an $I_{SPTA}$ of 70 mW/cm$^2$ for 20 min for continuous stimulation for 7 days. The cells were fixed with 4% PFA at 25 °C for 30 min, permeabilized for 10 min, and incubated with the primary antibody overnight at 4 °C. Then, the cells were incubated with the corresponding secondary antibodies for 45 min, washed three times with PBS, and incubated with DAPI for 15 min. The following antibodies were used: TNF-α, SOX-9, and COL-II which were purchased from Abcam. Goat anti-rabbit and goat anti-mouse secondary antibodies were purchased from Millipore (Billerica, MA, USA).

### RNA sequencing analysis

For the transcriptome study, the cells were divided into three groups: Ctrl, US-70, and US-100 stimulated with an $I_{SPTA}$ of 0, 70, or 100 mW/cm$^2$ LIPUS for 20 min daily. RNA samples were collected by TRIzol after 7 days of treatment and then stored in liquid nitrogen before being sent to the Beijing Genomics Institute for RNA sequencing analysis. After obtaining the sequencing data, the data were quantitatively analyzed using principal component analysis, differentially expressed gene (DEG) screening, GO function significant enrichment analysis, and pathway significant enrichment analysis.

### Animal surgery

Sprague–Dawley (SD) rats were purchased from Tongji University; all procedures of animal usage were approved by the Ethical Committee of Tongji Hospital affiliated to Tongji University (ID: [2019] GXB-01). In this study, 24 male SD rats (10 weeks old) were randomly divided into the following groups: Ctrl, LIPUS, hUC-MSCs, and LIPUS & hUC-MSCs. In the surgical preparation, the same classical rat knee cartilage defect model [47–50] was constructed for each group. The instruments were sterilized before the operation, and all animal-related operations were performed on a sterile professional surgical operating table. The SD rats were anesthetized, and the skin was removed to expose the knee joints of the lower limbs of the rat. The skin and muscle layers were incised on the inner side of the knee joint of rats. After exposing the knee joint, a full-thickness cylindrical cartilage defect with a diameter of 2 mm and a depth of 1.5 mm was drilled using a micro bone drill. After the defect was prepared successfully, the surgical wound was rinsed with sterile saline. For the Ctrl and LIPUS groups, the externally dislocated patella was repositioned and double sutured, while the joint capsule and wound were interrupted with 5.0 silk sutures, respectively. For the hUC-MSCs and LIPUS & hUC-MSCs groups, in vivo cell transplantation was performed prior to being sutured. A total of $1 \times 10^6$ hUC-MSCs/rat were counted and placed in a 20-μL chondrogenic medium; the cells were slowly injected into the articular cavities in the two side knees using a microinjector, 10 μL, respectively, and the defect area was sutured after resting 5 min later for cell stabilization. Then, the externally dislocated patella was repositioned and double sutured, while the joint capsule and wound were interrupted with 5.0 silk sutures, respectively.

The rats were anesthetized with 1.5% isoflurane and then placed in a suitable restraint board with head immobilized. The base of the skull was grasped firmly with one hand, while the other hand was used to apply a sharp and quick force to the neck, dislocating the cervical vertebrae and severing the spinal cord. The rats were closely monitored for cessation of breathing and heartbeat to ensure that they were euthanized effectively and humanely. The personnel performing the cervical dislocation method were trained and experienced, and we ensured that the procedure was carried out with the utmost care and compassion for the animals.

### Histological analysis

Knee joint specimens were collected after 2, 4, and 6 weeks. The samples were fixed with 4% PFA on a refrigerator shaker at 4 °C for 48 h, and the liquid was changed halfway. Each sample was decalcified with ethylenediaminetetraacetic acid (EDTA) decalcification solution (replaced every 3–5 days) and the decalcification effect was identified once a month until the knee joint specimen could be easily pierced by a needle. After decalcification was completed, the processes of grade ethanol dehydration, xylene permeation, and paraffin embedding were performed. The section thickness was 5 mm and stained with Safranin O-fast green and H&E staining.

### Immunochemistry staining

The sections of the knee joint were conducted with immunochemistry staining of COL-II, TNF-α, and CD44. The sections were incubated with pepsin antigen repair solution (0.4%) after dewaxing and hydration in xylene and gradient ethanol solution, permeabilized and blocked with blocking buffer (0.1% Triton X-100, 0.05% Tween-20, and 10% normal goat serum in PBS) for 1 h at room temperature before primary antibody incubation. The following primary antibodies were used: COL-II (Abcam), TNF-α (Abcam), and CD44 (CST). The sections were stained with responsive secondary antibodies, counterstained with DAB, and mounted with Fluoromount-G anti-fade mounting medium (Southern Biotechnology). Tissue sections were imaged using an Olympus microscope.

### Statistical analysis

One-way or two-way analysis of variance (ANOVA) was applied using GraphPad Prism software (v7.0) for all statistical analyses. The statistical analysis in this study was performed with at least three independent repetitions. Data are reported as means with standard deviation; $^*p<0.05$, $^{**}p<0.01$, and $^{***}p<0.001$ were considered statistically significant.

## Results

### Identification of hUC-MSCs and the effect of LIPUS on the proliferation of hUC-MSCs

The six-channel ultrasound stimulation system and the process of in vitro LIPUS stimulation are shown in Fig. 1a, where the transducers were placed on the porous foam cardboard to absorb the back-radiated ultrasound waves. The parameters of in vitro LIPUS stimulation are listed in Materials and Methods, with $I_{SPTA}$ ranging from 30 to 250 mW/cm$^2$ (Fig. 1b). For hUC-MSCs, we first identified specific markers on the cell surface, where CD29, CD44, CD73, CD105, and CD166 were positive markers and CD11b, CD14, CD34, CD45, and HLA-DR were negative markers (Fig. 1d). The hUC-MSCs formed a slender fusiform shape under the microscope and were successfully induced to differentiate in three directions: adipogenesis, chondrogenesis, and osteogenesis (Fig. 1c). To select suitable conditions for LIPUS stimulation, we set up eight different groups: Ctrl, US-30, US-50, US-70, US-100, US-150, US-200, and US-250, with respective $I_{SPTA}$ of 0, 30, 50, 70, 100, 150, 200, and 250 mW/cm$^2$. Compared with Ctrl group, the apoptosis rate of hUC-MSCs did not change significantly with LIPUS stimulation, except for the apoptotic rates of US-200 and US-250 groups which increased slightly at 9.5% and 11.3%, respectively (Fig. 1e, 1h). In addition, the results of the cell cycle showed that as the sound intensity of LIPUS increased, the G2 phase of the cell cycle also increased, while the corresponding S phase decreased. The most obvious changes were observed in US-200 and US-250 groups (Fig. 1f, 1i). The results of live–dead cell staining (Fig. 1g) also showed that the number of live cells was significantly decreased in US-200 and US-250 groups compared with Ctrl group.

### LIPUS promotes the chondrogenic differentiation of hUC-MSCs

To further identify the effects of LIPUS stimulation on chondrogenic differentiation of hUC-MSCs, we set up six groups based on the above-mentioned cell proliferation results: Ctrl, US-30, US-50, US-70, US-100, and US-150, excluding the higher sound intensity groups, US-200 and US-250, which showed negative effects on cell proliferation based on the above-mentioned results. After 7 continuous days of 20 min/day LIPUS stimulation, Alcian blue staining of hUC-MSCs in each LIPUS stimulation group was more positive than that of Ctrl group (Fig. 2a). The quantitative data showed a trend of increasing proteoglycan production with increasing sound intensity; further, the proteoglycan production was significantly higher under LIPUS stimulation with the sound intensity $I_{SPTA}$ of 70–150 mW/cm$^2$ than the Ctrl group (Fig. 2b). This was also confirmed via immunofluorescence staining of cytoskeleton dye actin and the two marker genes of chondrogenic differentiation, SOX-9 and ACAN. Staining of the cytoskeleton revealed that the actin of US-70 and US-100 groups had the highest brightness, and the cell morphology was also significantly broadened, which gradually differentiated into mature chondrocyte morphology (Fig. 2c). Meanwhile, the expression levels of *SOX-9* and *ACAN* significantly increased in US-70 and US-100 groups (Fig. 2d–e). In addition, qPCR detection and analysis showed that the hUC-MSCs treated with LIPUS stimulation could significantly up-regulate the expression of marker genes in the process of cartilage formation, such as *SOX-9*, *COL-II*, *COL-X*, *ACAN*, and *COMP*, of which US-70 group had the highest upregulation level (Fig. 2f). Compared with Ctrl group, the relative fold-change expression of *SOX-9*, *COL-II*, *COL-X*, *ACAN*, and *COMP* in US-70 group was 1.628, 2.357, 1.843, 2.292, and 1.689, respectively.

### LIPUS promotes anti-inflammation in C28/I2 cells

To explore the effect of LIPUS stimulation on the proliferation and activity of human mature chondrocytes, we set up six groups as aforementioned: Ctrl, US-30, US-50, US-70, US-100, and US-150. After two continuous days of 20 min/day LIPUS stimulation, the effect of LIPUS on cell cycle phases and apoptosis of C28/I2 cells was detected using flow cytometry. The results showed that there was no significant difference in the apoptotic rate (Fig. 3a–b) and cell cycle phase (Fig. 3c–d) for each LIPUS stimulation group compared with Ctrl group. The results of live–dead cell staining (Fig. 3e) also showed that the number of live cells showed little difference between the six groups.

To further study the effect of LIPUS stimulation on human chondrocytes in an inflammatory state, we set up a total of 12 groups with six groups as mentioned above: Ctrl, US-30, US-50, US-70, US-100, and US-150, whereas the other six groups were pretreated with 5 μg/mL lipopolysaccharide (LPS) to induce inflammation in chondrocytes; these groups were: LPS and LPS with either US-30, US-50, US-70, US-100, or US-150. After two days of 20 min/day LIPUS stimulation, the expression of signature genes associated with C28/I2 inflammation was detected via qPCR (Fig. 3f). The results of LPS and Ctrl groups showed that the expression of inflammation-related genes, including *TNF-α*, *IL-1β*, and *IL-6*, significantly increased after LPS treatment, with fold-changes of 3.492, 9.034, and 24.85, respectively. However, the results of LIPUS and LPS & LIPUS groups showed that the expression of these inflammatory marker genes was generally lower compared with those in Ctrl and LPS groups, respectively.

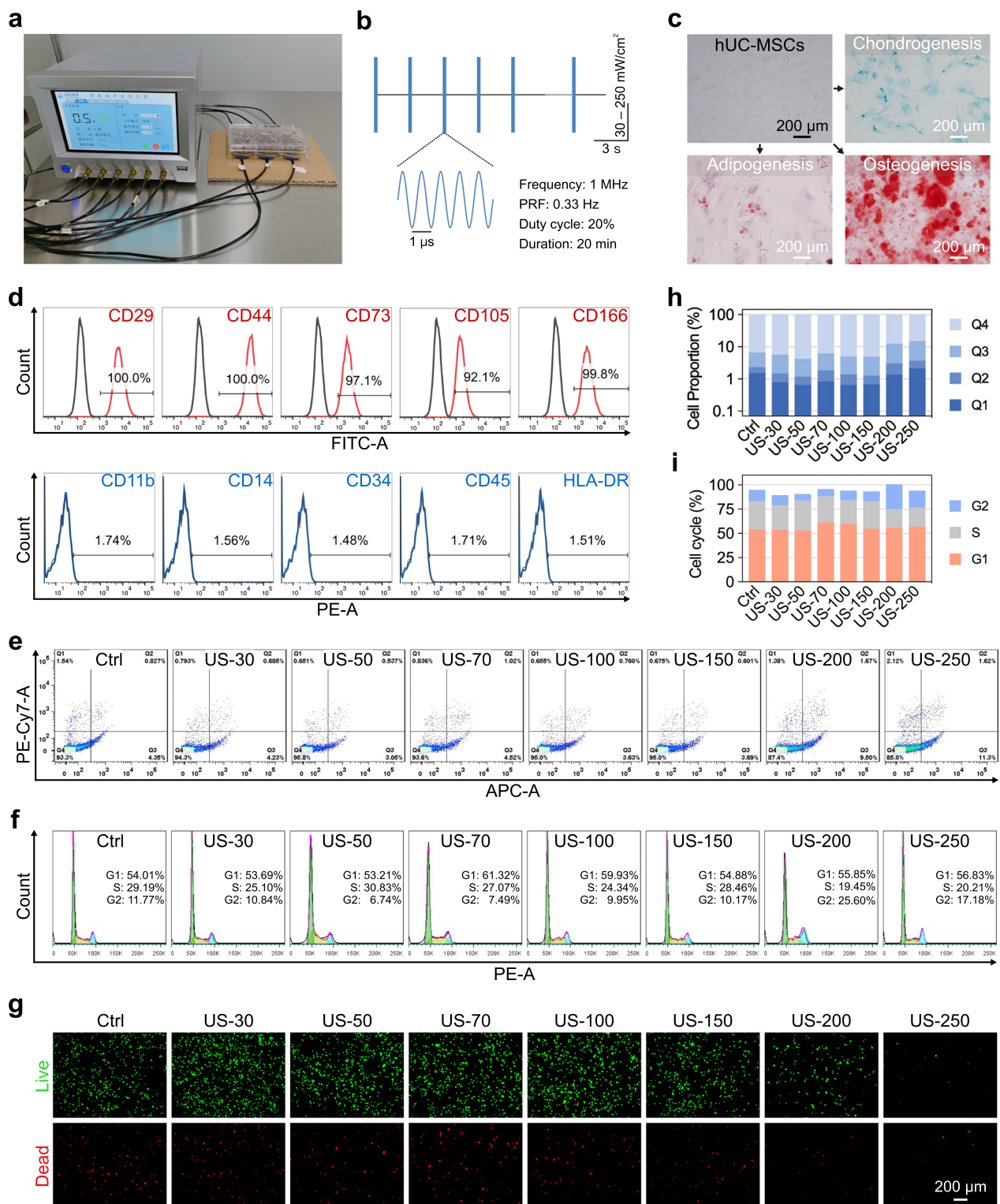

**Fig. 1** Identification of hUC-MSCs and the effect of LIPUS on the proliferation of hUC-MSCs. **a** The six-channel ultrasound stimulation system used for in vitro LIPUS stimulation. **b** LIPUS parameters. **c** The tri-lineage differentiation capability identification of hUC-MSCs; scale bar: 200 μm. **d** Flow cytometric identification of specific surface markers of hUC-MSCs. **e** Flow cytometry analysis for cell apoptosis of hUC-MSCs stimulated with LIPUS. **f** Flow cytometry analysis for cell cycle distribution of hUC-MSCs stimulated with LIPUS. **g** Representative live-dead staining images of hUC-MSCs stimulated with LIPUS; scale bar: 200 μm. **h** Quantification data for apoptosis rate of hUC-MSCs stimulated with LIPUS. **i** Quantification data for the cell cycle of hUC-MSCs stimulated with LIPUS

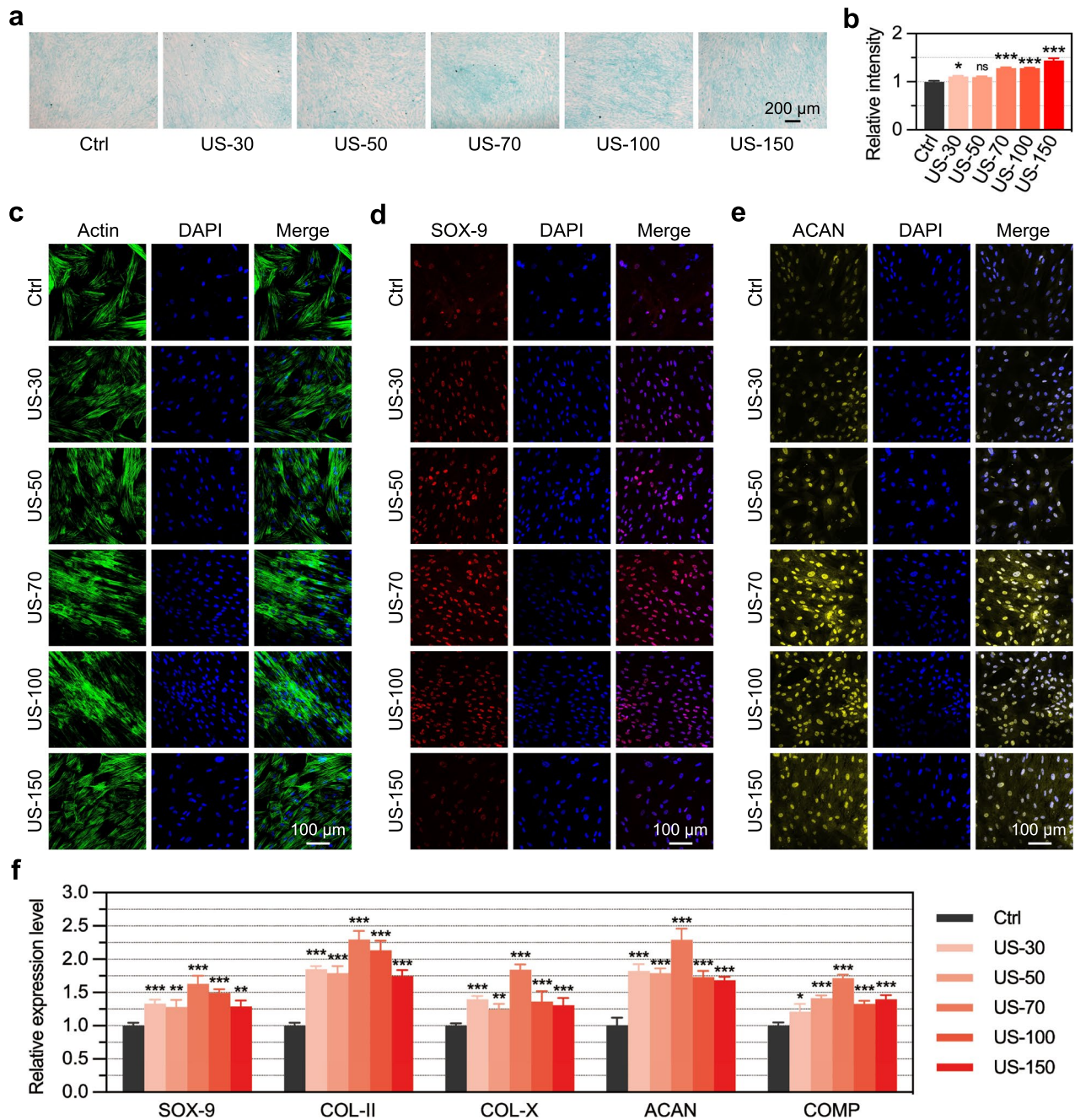

**Fig. 2** Effects of LIPUS on the chondrogenic differentiation of hUC-MSCs. The hUC-MSCs were stimulated by 20 min/day LIPUS stimulation for 7 days before analyses. **a** Representative Alcian blue staining images showing the differentiation level of hUC-MSCs into chondrocytes; scale bar: 200 μm. **b** Quantification data of the proteoglycan production for the Alcian blue staining results. **c–e** Representative immunofluorescence images showing the expression of cytoskeleton actin (**c**), and two marker genes of chondrogenic differentiation, SOX-9 (**d**) and ACAN (**e**); scale bar: 100 μm. **f** qPCR analysis of the expression of *SOX-9*, *COL-II*, *COL-X*, *ACAN*, and *COMP*. $*p < 0.05$, $**p < 0.01$, $***p < 0.001$ (n = 3)

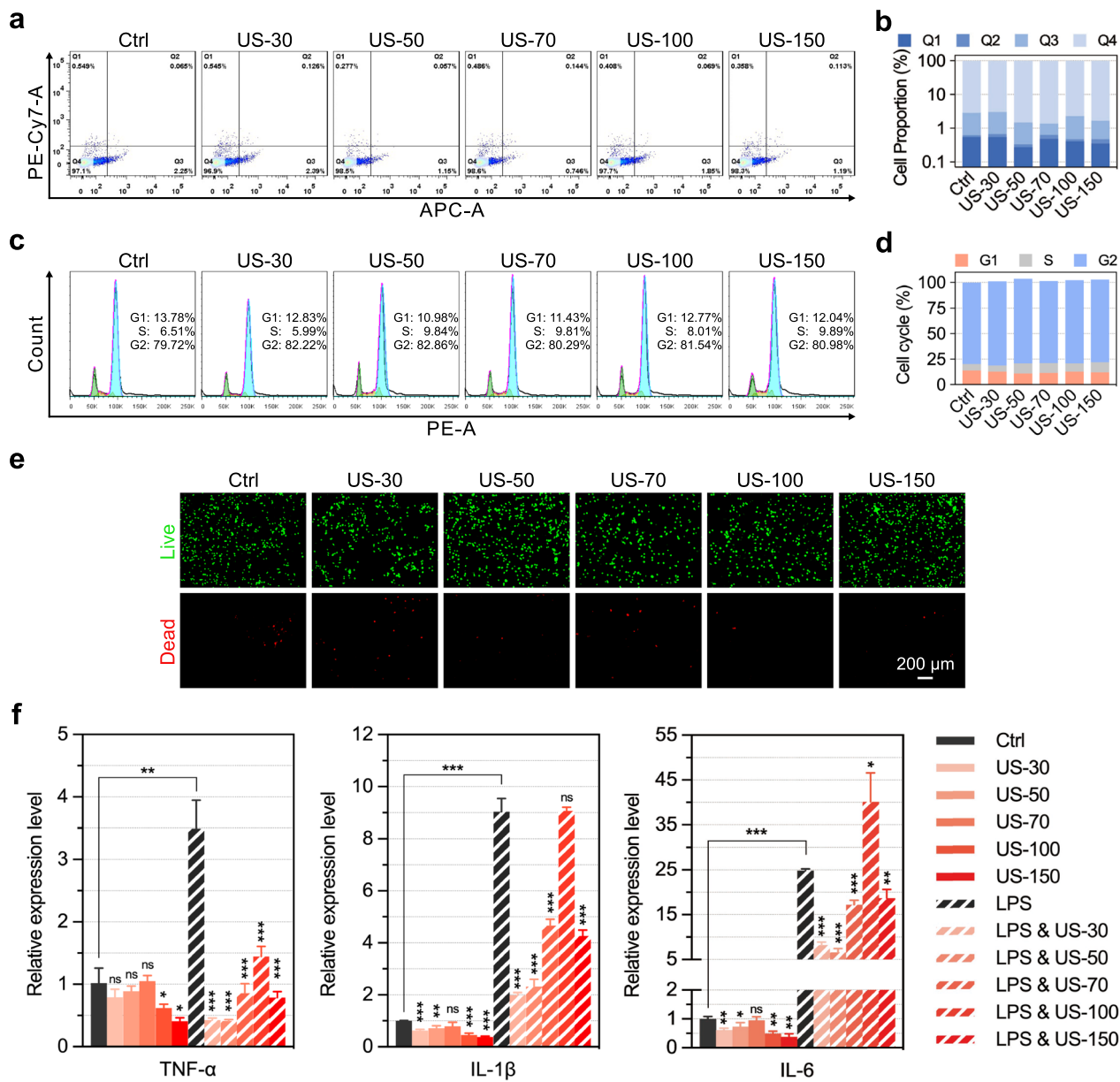

**Fig. 3** Effects of LIPUS stimulation on the proliferation and inflammation of C28/I2 cells. **a** Flow cytometry analysis for cell apoptosis of C28/I2 cells stimulated with LIPUS. **b** Quantification data for apoptosis rate of C28/I2 cells. **c** Flow cytometry analysis for cell cycle distribution of C28/I2 cells. **d** Quantification data for the cell cycle of C28/I2 cells. **e** Representative live-dead staining images of C28/I2 stimulated with LIPUS; scale bar: 200 μm. **f** qPCR analysis of the expression of *TNF-α*, *IL-1β*, and *IL-6* in C28/I2 cells, after LPS treatment (to produce inflammation) and LIPUS stimulation. $*p < 0.05$, $**p < 0.01$, $***p < 0.001$ (n = 3)

Particularly, the expression of TNF-α was significantly down-regulated in all LIPUS treatment groups, including US-70 group with an obvious promotion effect on the chondrogenic differentiation of hUC-MSCs, whose relative expression level dropped from 3.492 to 0.8572. This indicated that LIPUS stimulation resulted in the anti-inflammatory effect in C28/I2 cells.

### TNF signaling plays an important role in LIPUS stimulation-induced hUC-MSC chondrogenic differentiation

Our study showed that LIPUS with an $I_{SPTA}$ of either 70 or 100 mW/cm$^2$ could promote chondrogenic differentiation of hUC-MSCs. To further explore the potential molecular mechanism of LIPUS-induced chondrogenesis of hUC-MSCs, RNA sequencing was performed. The heatmap of the mRNA expression profile showed that the overall gene expression pattern of US-70 group was significantly different from that of the other two groups (Fig. 4a). Analysis of the differentially expressed genes (DEGs) between Ctrl and US-70 groups revealed that 1,673 genes were up-regulated and 476 genes were down-regulated (Fig. 4b). For all the up- and down-regulated DEGs, we performed Kyoto Encyclopedia of Genes and Genomes (KEGG) pathway term and Gene Ontology (GO) enrichment analyses. For the KEGG pathway term analysis, most genes were related to the organismal systems and human diseases (Fig. 4c). The KEGG analysis results revealed that most of the genes were involved in rheumatoid arthritis and TNF signaling pathways (Fig. 4d). Through GO enrichment analysis, we found that nuclear receptor activity, cytokine activity, and growth factor activity were captured by the related genes (Fig. 4e). To explore the molecular mechanism underlying LIPUS-induced MSC differentiation into cartilage from a more systematic perspective, we used a protein–protein interaction (PPI) network to analyze all related genes and found that the core regulatory genes in LIPUS that promote the cartilage differentiation of hUC-MSCs were *TNF-α*, *IL-6*, *CXCL8*, and *IL-1β* (Fig. 4f). In addition, by combining the results of the KEGG pathway term, KEGG, GO enrichment, and PPI analyses, we found that the TNF signaling pathway plays a leading role in the process of LIPUS-induced chondrogenic differentiation of hUC-MSCs.

Further research was performed to analyze the role of the TNF signaling pathway in LIPUS stimulation-induced hUC-MSC chondrogenic differentiation. Notably, the reduction in TNF-α levels after LIPUS stimulation plays a key role in the differentiation of hUC-MSCs into chondrocytes. First, we used three sequenced samples to detect the expression levels of the three inflammatory factor genes, *TNF-α*, *IL-1β*, and *CXCL8*. The results

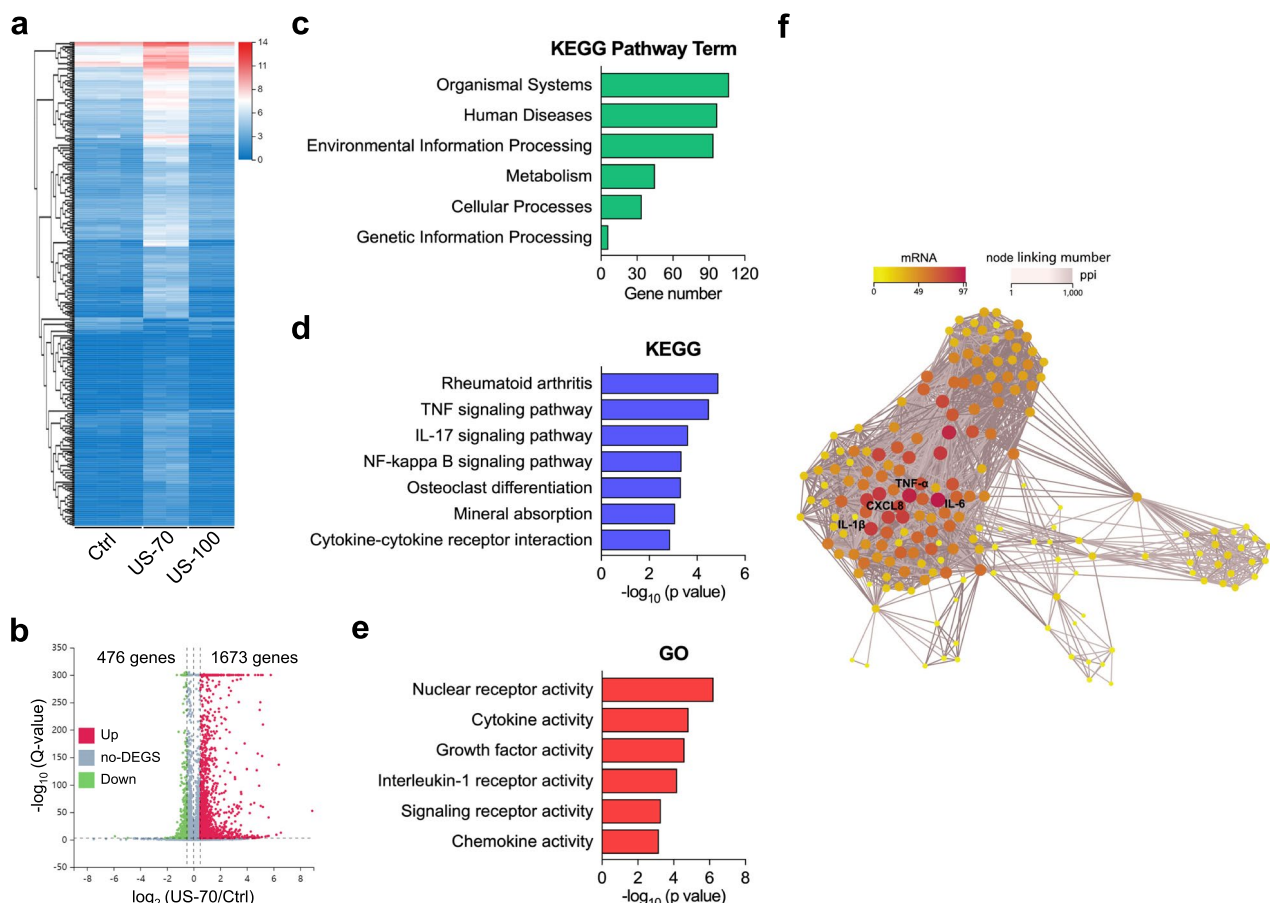

**Fig. 4** Transcriptome sequencing analysis. **a** Cluster heat map of mRNA expression of Ctrl, US-70, and US-100 samples. **b** A volcano plot comparing the DEGs between the US-70 and Ctrl groups (Down-regulated genes showed in green and up-regulated genes showed in red; cutoff: *p*-value < 0.05 and |log$_2$ FC|> 1). **c–e** KEGG pathway term analysis (**c**), KEGG pathway analysis (**d**), and GO pathway analysis of the DEGs (**e**) between the US-70 and Ctrl groups. **f** PPI network analysis of the core regulatory genes of the DEGs between the US-70 and Ctrl groups

showed that the expression levels of these inflammatory factors in US-70 group with the best chondrogenic differentiation effect were significantly decreased (Fig. 5a). Compared with Ctrl group, the relative expression of *TNF-α*, *IL-1β*, and *CXCL8* in US-70 group decreased to 0.1977, 0.7126, and 0.1197, respectively. Subsequently, to verify that LIPUS stimulation with an $I_{SPTA}$ of 70 mW/cm$^2$ can reduce the level of TNF-α, we set up four groups: Ctrl, LIPUS, TNF-α, and TNF-α & LIPUS, where the latter two groups were pretreated with 10 μg/mL TNF-α to induce inflammation in hUC-MSCs. The immunofluorescence staining results showed that the level of TNF-α was significantly lower in TNF-α & LIPUS group than in TNF-α group (Fig. 5b). The same was true when comparing LIPUS and Ctrl groups which did not have TNF-α-induced inflammation. Alcian blue staining results showed that the differentiation of hUC-MSCs into chondrocytes was inhibited following TNF-α upregulation (Fig. 5d). The quantitative statistical analysis of the proteoglycan production for the Alcian blue staining showed that, compared with the Ctrl group, TNF-α significantly reduced proteoglycan production, while LIPUS stimulation significantly increased proteoglycan production (Fig. 5c). To determine whether LIPUS stimulation with an $I_{SPTA}$ of 70 mW/cm$^2$ inhibited the expression of TNF-α and promoted the chondrogenic differentiation of hUC-MSCs, the cartilage marker genes *SOX-9* and *COL-II* were fluorescently labeled for immunofluorescence confocal imaging. The TNF-α & LIPUS group had better chondrogenic differentiation performance than TNF-α group, especially with a significantly increased brightness of SOX-9 and COL-II (Fig. 5e–f). In addition, these two marker genes of chondrogenic differentiation were weaker in TNF-α group than in Ctrl and LIPUS groups, which did not stimulate inflammation by TNF-α. These results indicated that LIPUS stimulation with an $I_{SPTA}$ of 70 mW/cm$^2$ could reduce the level of TNF-α in hUC-MSCs and promote hUC-MSC differentiation into chondrocytes.

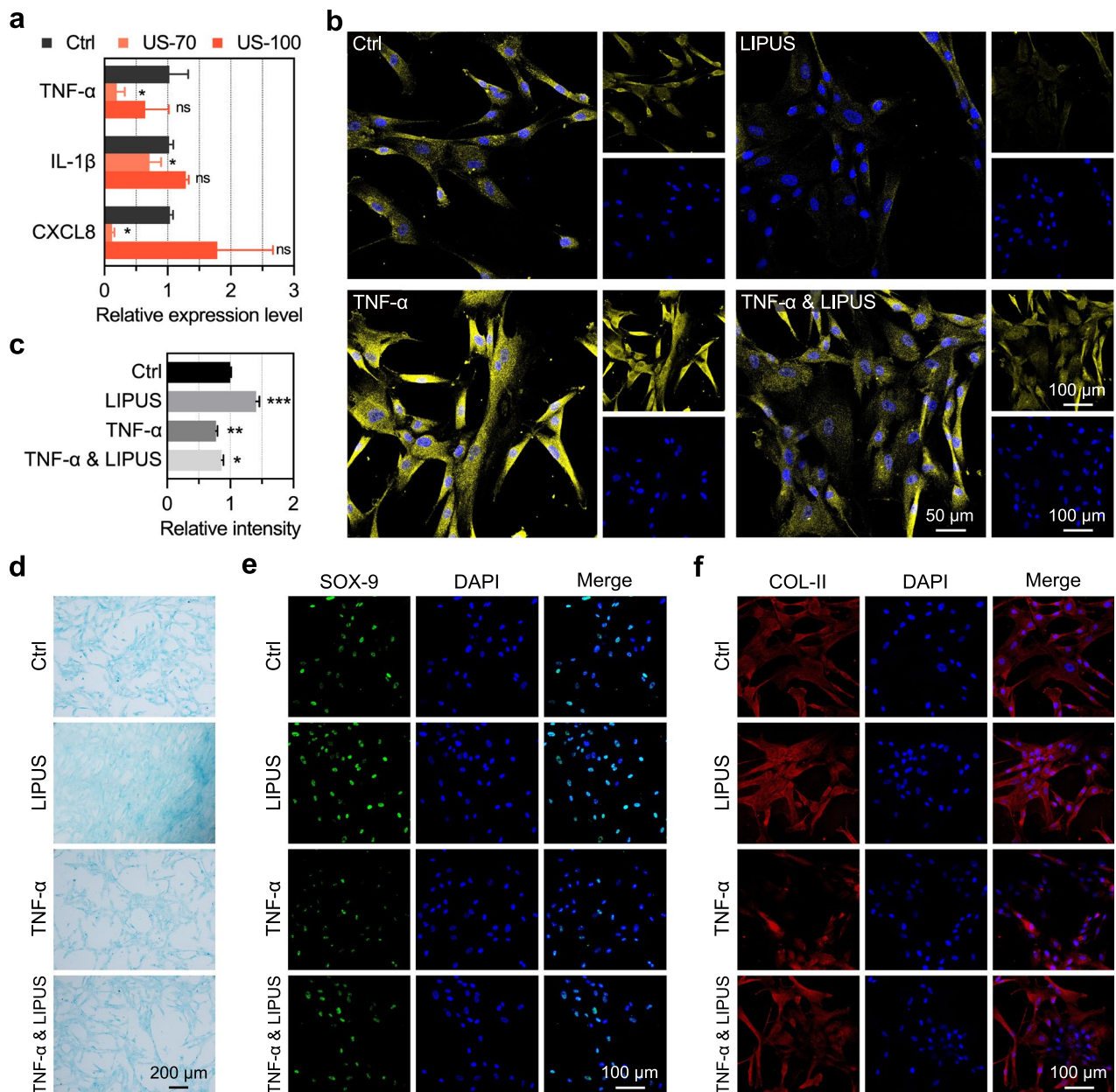

**Fig. 5** Effects of TNF signaling pathway in LIPUS stimulation-induced hUC-MSC chondrogenic differentiation. **a** qPCR analysis of the expression of inflammatory factor genes *TNF-α, IL-1β*, and *CXCL8*. *$p<0.05$, **$p<0.01$, ***$p<0.001$ (n = 3). **b** Representative immunofluorescence images showing the expression level of inflammatory factor TNF-α; scale bars: 50 μm and 100 μm, respectively. **c** Quantification data of the proteoglycan production for the Alcian blue staining results. **d** Representative Alcian blue staining images showing the differentiation level of hUC-MSCs into chondrocytes; scale bar: 200 μm. **e–f** Representative immunofluorescence images showing the expression of the chondrogenic differentiation markers SOX-9 (**e**) and COL-II (**f**); scale bar: 100 μm

**Validation of LIPUS- and hUC-MSC-based cartilage regeneration in vivo**

To verify the LIPUS- and hUC-MSC-based cartilage regeneration in vivo, rat knee articular cartilage defect models were established and divided into four groups: Ctrl (defect), hUC-MSCs, LIPUS, and LIPUS & hUC-MSCs. The in vivo LIPUS stimulation experimental setup is shown in Fig. 6a, where the rat was anesthetized, and the knee joint with a shaven surface was placed on the transducer. To estimate the ultrasonic attenuation through biological tissues, we simplified the complex propagation path from the ultrasound transducer to the cartilage defect which was divided into two layers of soft tissue and bone, each with a thickness of 3 mm. As the ultrasonic attenuation coefficients through soft tissue and bone are 1 dB/MHz/cm and 10 dB/MHz/cm, respectively [51], a total ultrasonic attenuation of approximately 3.3 dB was estimated, of which 0.3 dB is attributed to soft tissue and 3 dB to bone. Thus, the $I_{SPTA}$ of LIPUS stimulation for in vivo experiments was set to 150 mW/cm$^2$ with other LIPUS parameters the same as those used in in vitro stimulation. After continuous 20 min/day LIPUS stimulation for 2, 4, and 6 weeks, the knee joint samples were subjected to bright-field photography and histological staining with H&E and Safranin O-fast green to identify the healing effect of the treatment on the defective cartilage. The defective area of each group was observed, and no significant inflammation or synovitis was found on the slides. There were obvious signs of tissue regeneration at the defected area after 2 weeks in LIPUS & hUC-MSCs group, 6 weeks in hUC-MSCs and LIPUS groups with shallower defect depths, and almost no regenerative tissue in Ctrl group (Fig. 6b). The H&E and Safranin O staining of the defected model area (Fig. 6c) showed a small amount of regenerated tissue in Ctrl group, whereas more regenerated tissue was observed in the other three groups, especially LIPUS & hUC-MSCs group which showed the best recovery of the defected structure. In Ctrl and LIPUS groups, the new tissue stained with Safranin O was negative and did not well integrate with the surrounding healthy tissue, suggesting that the regenerated tissue was either not a cartilage tissue or still an immature tissue. In hUC-MSCs group, Safranin O staining was weakly positive in new tissues after 2 weeks, indicating that the regenerated tissue was an immature cartilage tissue, and the boundary between the injured and healthy areas was loosely defined. The new tissues in LIPUS & hUC-MSCs group were tightly connected to the surrounding healthy tissues, and Safranin O staining was positive after both 2 and 6 weeks, with almost no difference in the surrounding healthy tissues after 6 weeks. These results showed that the knee cartilage defect was significantly repaired in LIPUS & hUC-MSCs group compared with Ctrl group, and the repair cycle was shorter than that in hUC-MSCs and LIPUS groups. The in vivo immunohistochemistry results (Fig. 6d) of joint tissues at 6 weeks including collagen-II, TNF-α, and CD44 showed that cell transplantation (hUC-MSCs), ultrasound stimulation (LIPUS), and combined treatment (LIPUS & hUC-MSCs) groups could promote the expression of Collagen-II at 6 weeks, postoperatively compared to the Ctrl group; a significant number of CD44-positive cells were present in the hUC-MSCs and LIPUS & hUC-MSCs groups, while no such positive cells were present in the other groups, indicating that much MSCs were still alive 6 weeks after cell transplantation. Besides, after 6 weeks of postoperative treatment, both the Ctrl and hUC-MSCs groups elicited higher levels of TNF-α expression compared to the lower TNF-α expression levels in the Normal group, whereas the LIPUS and LIPUS & hUC-MSCs groups showed relatively lower TNF-α expression than Ctrl and hUC-MSCs groups. In addition, after 6 weeks of LIPUS treatment, the main organs (heart, liver, spleen, lung, and kidney) of the Sprague–Dawley (SD) rats in the different groups were stained with H&E. No noticeable tissue damage or pathological lesions were found in the organs of each group, which suggested the biosafety of LIPUS in vivo

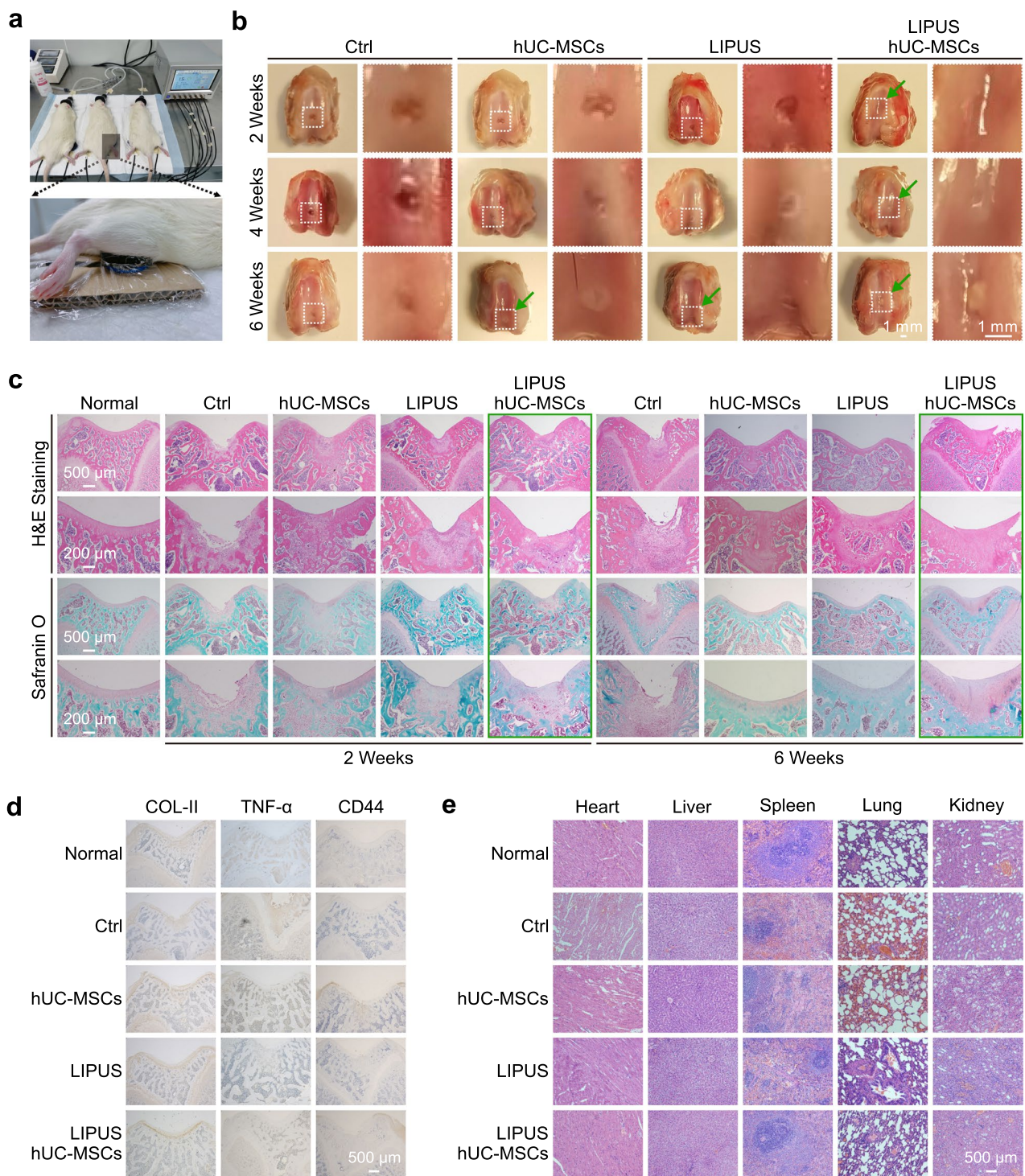

**Fig. 6** Validation of LIPUS and hUC-MSC-based cartilage regeneration in vivo. **a** Top—the ultrasound stimulation system for in vivo LIPUS stimulation on the rat knee joint. Bottom—partially enlarged photograph of the experiment setup. **b** Bright-field photography showing the appearance of rat knee joint samples after LIPUS stimulation for 2, 4, and 6 weeks; scale bar: 1 mm. In each group, left—overall view of the knee joint, right—partially enlarged photograph of the defect area. **c** Respective H&E staining and Safranin O-fast green staining of the paraffin sections of the knee joint defect showing the healing effect of LIPUS stimulation; scale bars: 500 μm for H&E staining and 200 μm for Safranin O-fast green staining, respectively. **d** Representative immunohistochemistry images of COL-II, TNF-α, and CD44 of the knee joint defect in rats after 6 weeks of LIPUS stimulation; scale bar: 500 μm. **e** Representative H&E staining images of major organs in rats after 6 weeks of LIPUS stimulation; scale bar: 500 μm

(Fig. 6e). Taken together, LIPUS stimulation could reduce the level of TNF-α in hUC-MSCs, promote hUC-MSC differentiation into chondrocytes, and thus realize articular cartilage regeneration (Fig. 7).

## Discussion

Low-intensity pulsed ultrasound (LIPUS), a form of ultrasound delivers at a low intensity and outputs in the mode of pulsed waves, has been demonstrated as adjuvant physical therapy, including promoting the differentiation of stem cells [22, 23], inhibiting inflammatory responses [24], accelerating soft tissue regeneration [25, 26], modulating neuronal activity [27, 28], improving bone healing [29, 30], and cartilage regeneration [31–35]. In this study, we investigated the promoting effects and mechanisms underlying LIPUS stimulation on the chondrogenic differentiation of human umbilical cord mesenchymal stem cells (hUC-MSCs) and further evaluated its regenerative application value in articular cartilage defects in rats.

Initially, we showed that LIPUS stimulation promotes hUC-MSC chondrogenic differentiation. The limited intrinsic healing ability of the cartilage hinders its ability to repair itself once damaged, requiring effective external treatment, whereas currently reported cartilage repair procedures are either invasive or have a limited repair-promoting ability [3–6]. Stem cell transplantation is a promising therapy for cartilage defect repair [11, 12], especially the rich sources, and chondrogenic differentiation ability of MSCs makes them an ideal cell type for cell transplantation-based cartilage tissue repair [14, 16–18]. Our initial in vitro analysis showed that LIPUS stimulation of hUC-MSCs up-regulated the expression of mature chondrocytes (Fig. 2a, Additional file 2: Fig. S1) and related marker genes such as *SOX-9* and *ACAN* (Fig. 2d–e, Additional file 2: Fig. S2), which indicated that LIPUS stimulation promotes the chondrogenic differentiation of hUC-MSCs and has the potential to facilitate the regeneration of damaged articular cartilage. In particular, LIPUS with an $I_{SPTA}$ of 70 mW/cm$^2$ showed the most significant promotion effect.

Secondly, our results suggested that TNF signaling pathway inhibition is a key factor in LIPUS-induced hUC-MSC chondrogenic differentiation. To date, numerous studies have suggested various possible mechanisms by which LIPUS promotes MSC chondrogenesis, including regulation of autophagy [43] and upregulating the expression of cyclin-D1 [40]. Previous research has reported that biophysical stimulation can activate the MARK and PI3K-Akt signaling pathways and promote cell proliferation and differentiation [52]. However, its specific mechanism remains unclear. In this study, RNA sequencing was performed to further explore the potential molecular mechanism of LIPUS-induced chondrogenesis of hUC-MSCs, with the TNF signaling pathway gene expression found to have significant changes after LIPUS stimulation with an $I_{SPTA}$ of 70 mW/cm$^2$ (Fig. 4d). We further verified that LIPUS stimulation with an $I_{SPTA}$ of 70 mW/cm$^2$ can significantly down-regulate the expression levels of *TNF-α* (Fig. 5a), reduce inflammation of hUC-MSCs induced by TNF-α (Fig. 5b), improve the level of chondrogenic differentiation of hUC-MSCs (decreased by TNF-α) (Fig. 5d), and up-regulate the expression of the cartilage marker genes *SOX-9* and *COL-II* (down-regulated by TNF-α) (Fig. 5e–f). The results indicated that the inhibition of TNF signaling pathway, especially *TNF-α* gene expression downregulation, plays a key role in LIPUS-induced hUC-MSC chondrogenic differentiation.

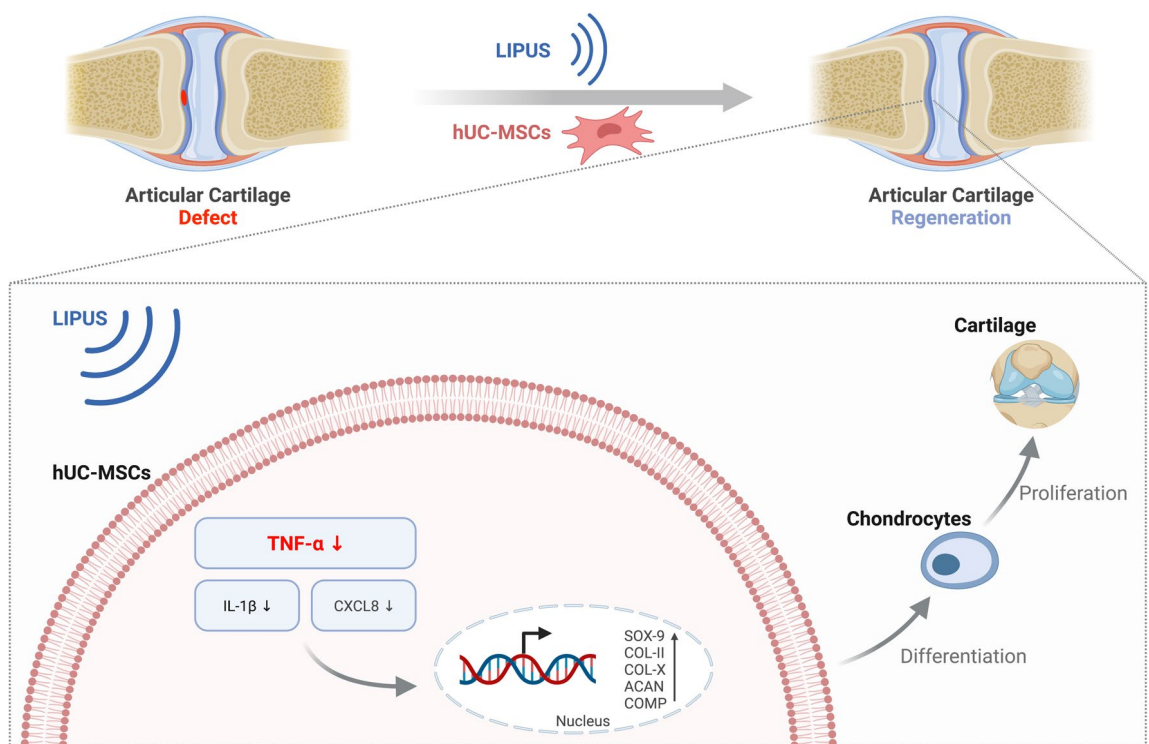

**Fig. 7** Schematic representation of LIPUS stimulation and hUC-MSC transplantation-based articular cartilage regeneration. Created with BioRender.com

Thirdly, we demonstrated that LIPUS provided a favorable anti-inflammatory microenvironment for newly differentiated chondrocytes and exhibited an anti-inflammatory effect in mature chondrocytes. The cartilage remodeling process is mediated by chondrocytes, whereas TNF-α has been reported to drive the inflammatory process in chondrocytes [53]. Besides, previous studies have reported that the levels of pro-inflammatory cytokines, including *TNF-α* and *IL-1β*, were markedly increased in patients with OA compared to healthy populations, which could lead to the damage of the articular cartilage [54]. High concentrations of *IL-1β* could up-regulate the expression of *MMP-1*, *MMP-3*, and *MMP-13* in chondrocytes, inhibit the expression of *ACAN* and *COL-II* with cartilage properties, and promote the production of *COL-I* with fibroblast characteristics, thereby destroying cartilage stability [55].

Effective reduction in the concentration of *IL-1β* in the cartilage tissue by antagonizing *IL-1β* treatment can simultaneously reduce the expression of *TNF-α* and *IL-8*, which could slow down cartilage damage and OA development. Nakao et al. found that LIPUS inhibits LPS-induced inflammatory chemokine mRNA expression in osteoblasts, including that of *CXCL1* and *CXCL10* [24]. In this study, mature chondrocytes, C28/I2 cells, were pretreated with LPS to induce inflammation and simulate an inflammatory state. The results showed that the expression of *TNF-α*, *IL-1β*, and *IL-6* significantly increased after LPS treatment, yet decreased after LIPUS stimulation (Fig. 3f), indicating that LIPUS stimulation exhibited an anti-inflammatory effect in mature chondrocytes. In addition, LIPUS stimulation significantly down-regulated the expression levels of inflammatory factors including *TNF-α*, *IL-1β*, and *CXCL8* in hUC-MSCs (Fig. 5a), showing that LIPUS stimulation provided a favorable anti-inflammatory microenvironment for newly differentiated chondrocytes. Thus, the anti-inflammatory effect of LIPUS can help to promote hUC-MSC transplantation-based articular cartilage regeneration or can be further used for OA treatment.

Lastly, we established a rat knee articular cartilage defect model to verify LIPUS- and hUC-MSC-based cartilage regeneration in vivo. The bright-field photography (Fig. 6b) and histological staining results (Fig. 6c) indicated that the knee cartilage defect was significantly repaired in LIPUS & hUC-MSCs group compared with Ctrl group, and the repair cycle was shorter than that in the respective hUC-MSCs and LIPUS groups. Our results suggest that LIPUS stimulation not only has therapeutic potential in treating articular cartilage injury but can also promote hUC-MSCs transplantation-based cartilage regeneration treatment.

Notably, LIPUS group and hUC-MSCs group had similar cartilage repair effects, both having shallower defect depths at week 6 compared to the Ctrl group, which indicated that LIPUS stimulation (without hUC-MSC transplantation) contributes to articular cartilage regeneration. In addition, the immunostaining results (Fig. 6d) showed that the LIPUS and LIPUS & hUC-MSCs groups showed relatively lower TNF-α expression than the Ctrl and hUC-MSCs groups, which indicated that ultrasound stimulation can regulate the expression of TNF-α and thus function in cartilage regeneration. Therefore, we suggest that the repair effect of LIPUS stimulation itself may be associated with the anti-inflammatory effect of LIPUS on chondrocytes.

In addition, the origin of the newly formed chondrocytes in the defect region must be determined and may involve the differentiation of exogenous (transplanted hUC-MSCs) or endogenous stem cells [56, 57]. However, cartilage tissue lacks endogenous stem cells and blood vessels for self-repair [2]. In this study, considering the correlation between transplanted hUC-MSCs and newly formed cartilage tissue, immunohistochemical staining for human anti-CD44 antibody was performed on the samples to effectively label the transplanted human MSCs [58]. Only human MSCs would be labeled by the anti-CD44 antibody, which were not present in the original rat model. The results showed that the CD44-positive cell markers were present in the cell transplantation involving groups including the hUC-MSC and LIPUS & hUC-MSC groups, at 6 weeks postoperatively (Fig. 6d), whereas such positive cells were not present in the other groups, indicating that CD44 effectively labeled hUC-MSCs that transplanted into the defect region. In addition, immunohistochemical staining results for the inflammatory marker TNF-α (Fig. 6d) showed that after 6 weeks treatment, the ultrasound stimulation involving groups including the LIPUS and LIPUS & hUC-MSCs groups exhibited relatively lower TNF-α expression, i.e., relatively lower levels of inflammation, compared with the Ctrl and hUC-MSCs groups, indicating that LIPUS stimulation had some anti-inflammatory effect. Moreover, the actual cartilage repair effect was evaluated via bright-field photography (Fig. 6b) and histological staining results (Fig. 6c), and the LIPUS & hUC-MSCs groups showed significant repair of the knee cartilage defects compared to the other three groups. In conclusion, treatment of the LIPUS & hUC-MSC groups enabled better survival of the transplanted cells in the inflammatory microenvironment to resist inflammatory stimuli in the defect region, thus effectively differentiating into cartilage and promoting tissue regeneration and repair.

Furthermore, we verified the safety of LIPUS stimulation and hUC-MSC transplantation-based cartilage repairing procedure. LIPUS non-invasively delivers mechanical stimulation to biological tissues without thermal activity and has been extensively exploited as a therapeutic tool in tissue engineering research, including cartilage regeneration [31–35]. However, it remains vital to determine and validate the LIPUS parameters for its safe and effective use. Previous studies that showed effective LIPUS stimulation on cartilage regeneration [59], bone healing [60], and dentofacial tissue engineering [61, 62] commonly used a duty cycle of 20%, 20 min/day stimulation, center frequency from 1 to 3 MHz, and $I_{SPTA}$ from 30 to 300 mW/cm$^2$. In this study, the parameters of LIPUS stimulation were as follows: center frequency = 1 MHz, pulse repetition frequency = 0.33 Hz, duty cycle = 20%, and $I_{SPTA}$ ranging from 30 to 250 mW/cm$^2$. The LIPUS parameters used in our study were first tested in vitro to ensure the security for the subsequent in vivo analysis. We investigated the cell cycle phases and apoptosis (to reflect the vitality of cells and therefore LIPUS safety) and found that these parameters were significantly changed under LIPUS stimulation with an $I_{SPTA}$ of 200 and 250 mW/cm$^2$ (Fig. 1e–g), indicating that LIPUS stimulation with an $I_{SPTA}$ of over 200 mW/cm$^2$ is not conducive to cell growth. Therefore, we conducted further research in the range of 30–150 mW/cm$^2$.

We performed tissue biosafety evaluation using H&E staining of major organs, including the heart, liver, spleen, lung, and kidney, from the rats after 6 weeks of treatment (Fig. 6e), where no noticeable tissue damage or pathological lesions were found in any of the groups, indicating the safe application of the LIPUS stimulation. Thus, the safety and effectiveness of LIPUS stimulation parameters were verified both in vitro and in vivo in this study. In addition, previous research has suggested that hUC-MSCs exhibit lower immunogenicity compared with other types of MSCs [63–65]. Hence, we verified the safety of LIPUS stimulation and hUC-MSC transplantation-based cartilage repairing.

LIPUS has been extensively studies in tissue engineering research, including cartilage regeneration. For example, Xia et al. found that LIPUS stimulation promoted TGF-β1-induced chondrogenesis of bone marrow mesenchymal stem cells (BMSCs) through the integrin-mTOR signaling pathway [66], while Wang et al. found that chondrogenesis of BMSCs could be promoted by LIPUS via regulation of autophagy [43]. Ling et al. found that LIPUS pre-treatment can improve the effect of hAD-MSC transplantation by downregulating the expression of pro-inflammatory cytokines (IL-1β, IL-6, and TNF-α) and reducing ovarian inflammation induced by chemotherapy [22]. Furthermore, Liu et al. found that LIPUS stimulation inhibited the expression of inflammatory factors (IL-6, IL-8) and promoted the osteogenic differentiation capacity of human periodontal ligament cells (hPDLCs) by inhibiting the NF-κB signaling pathway [67]. Xia et al. constructed a rat knee osteoarthritis (OA) model and demonstrated that LIPUS can enhance the therapeutic efficacy of BMSCs in OA cartilage repair, and the underlying mechanism is associated with the increase in autophagy-mediated exosome release [68]. In the current study, LIPUS stimulation with specific parameters promoted chondrogenic differentiation and down-regulated the expression levels of inflammatory factors TNF-α, IL-1β, and CXCL8 in hUC-MSCs, which are associated with the TNF signaling pathway inhibition. Although the concept of "LIPUS-stimulated and hUC-MSC-mediated cartilage regeneration" has been reported in several studies, the present study validated the promoting effect of LIPUS on the chondrogenic differentiation of hUC-MSCs in injured articular cartilage rat models in vivo, with a new possible regenerative mechanism related to LIPUS-induced TNF signaling pathway inhibition. This study elucidates the possible mechanistic basis for LIPUS-stimulated and hUC-MSC-mediated cartilage regeneration, offering a practical and bio-safe procedure for articular cartilage repair and regeneration for clinical applications.

Nevertheless, this study has some limitations. First, the ultrasound transducers we used convey plane waves but not focused waves that could focus ultrasound energy on the target cartilage defect area, which may limit the stimulation efficiency. Second, the impact of other LIPUS parameters including central frequency and duty cycle was not studied. Third, only the TNF signal pathway was studied, among differential gene pathways following LIPUS stimulation, which may also be associated with the mechanism underlying LIPUS-induced hUC-MSC chondrogenic differentiation. In future research, we will use focused ultrasound transducers to further study the impact of LIPUS on other differential gene pathways and try to verify the hybrid repairing procedure using animals with cartilages closer to the human articular cartilage.

## Conclusion

Our data show that LIPUS stimulation with specific parameters can promote chondrogenic differentiation of hUC-MSCs by inhibiting the TNF signaling pathway and provide a favorable anti-inflammatory microenvironment for the newly differentiated chondrocytes, thus promoting hUC-MSC transplantation-based articular cartilage regeneration in vivo. The present study elucidates the mechanistic basis for cartilage regeneration based on hUC-MSC transplantation and LIPUS stimulation, thus offering an effective and bio-safe procedure for articular cartilage repair and regeneration for clinical applications.

### Abbreviations

| | |
|---|---|
| ACAN | Aggrecan |
| ACI | Autologous chondrocyte implantation |
| ANOVA | Analysis of variance |
| COL-I | Type I collagen |
| COL-II | Type II collagen |
| COL-X | Type X collagen |
| COMP | Cartilage oligomeric matrix protein |
| CXCL8 | C–X–C motif chemokine ligand 8 |
| DEG | Differentially expressed gene |
| FBS | Fetal bovine serum |
| H&E | Hematoxylin–eosin |
| hAD-MSCs | Human adipose-derived mesenchymal stem cells |
| hBMSCs | Human bone mesenchymal stem cells |
| hUC-MSCs | Human umbilical cord mesenchymal stem cells |
| hUCB-MSCs | Human umbilical cord blood-derived mesenchymal stem cells |
| IL-1β | Interleukin-1β |
| IL-6 | Interleukin 6 |
| $I_{SPTA}$ | Spatial-peak temporal-average intensity |
| LIPUS | Low-intensity pulsed ultrasound |
| LPS | Lipopolysaccharide |
| MSCs | Mesenchymal stem cells |
| OA | Osteoarthritis |
| P/S | Penicillin–streptomycin |
| PBS | Phosphate-buffered saline |
| PFA | Paraformaldehyde |
| PPI | Protein–protein interaction |
| SD | Sprague–Dawley |
| SOX-9 | Sex-determining region Y-box 9 |
| TNF | Tumor necrosis factor |
| TNF-α | Tumor necrosis factor-α |

### Supplementary Information

The online version contains supplementary material available at https://doi.org/10.1186/s13287-023-03296-6.

**Additional file 1 Table S1**: Lists of hUC-MSCs qPCR primers. **Table S2**: Lists of C28/I2 qPCR primers

**Additional file 2 Figure S1**: Representative Alcian blue staining images for the chondrocyte spheres in suspension culture for LIPUS stimulation and adhered for staining. **Figure S2**: Western blot data of chondrogenic proteins (COL-II, ACAN, and SOX-9) and actin after different parameters of LIPUS stimulation

**Additional file 3** Full-length blots of the data shown in Figure S2.

#### Acknowledgements
We would like to thank Editage (http://www.editage.cn) for English language editing of the manuscript.

#### Author contributions
QC and LC designed and conceived the study. YC and HY performed the experiments and wrote the manuscript. ZW edited and reviewed the manuscript. RZ provided experimental equipment. All authors have read and agreed to the published version of the manuscript.

#### Funding
This work was supported by the National Natural Science Foundation of China (12034015), National Key Research and Development Program of China (2016YFA0100800), Shanghai Municipal Science and Technology Major Project (2021SHZDZX0100), Shanghai Municipal Commission of Science and Technology Project (19511132101), and Program of Shanghai Academic Research Leader (21XD1403600).

#### Availability of data and materials
The datasets used and/or analyzed during the current study are available from the corresponding author on reasonable request. The raw sequence data reported in this paper have been deposited in the Genome Sequence Archive (Genomics, Proteomics & Bioinformatics 2021) in National Genomics Data Center (Nucleic Acids Res 2022), China National Center for Bioinformation/Beijing Institute of Genomics, Chinese Academy of Sciences (GSA-Human: HRA004058), that are publicly accessible at https://ngdc.cncb.ac.cn/gsa-human.

### Declarations

**Ethics approval and consent to participate**
All experiments were performed in accordance with the guidelines and ethical procedures of Tongji Hospital affiliated to Tongji University. All procedures of animal usage were approved by Shanghai Tongji Hospital Stem Cell Clinical Research Ethics Committee (title of the approved project: Highly selective human umbilical cord mesenchymal stem cells for the repair of cartilage defects; approval number: [2019] GXB-01; date of approval: April 11, 2019).

**Consent for publication**
Not applicable.

**Competing interests**
The authors declare no competing interests.

#### Author details
[1]Institute of Acoustics, School of Physics Science and Engineering, Tongji University, Shanghai 200092, China. [2]Key Laboratory of Spine and Spinal Cord Injury Repair and Regeneration of Ministry of Education, Department of Orthopedics, Tongji Hospital affiliated to Tongji University School of Medicine, Tongji University, Shanghai 200065, China. [3]Frontiers Science Center for Intelligent Autonomous Systems, Shanghai 201210, China. [4]School of Life Science and Technology, Tongji University, Shanghai 200065, China.

### References
1. Buckwalter JA. Articular cartilage injuries. Clin Orthop Relat Res. 2002;402:21–37.
2. Pearle AD, Warren RF, Rodeo SA. Basic science of articular cartilage and osteoarthritis. Clin Sports Med. 2005;24:1–12.
3. Kwon H, Brown WE, Lee CA, Wang D, Paschos N, Hu JC, et al. Surgical and tissue engineering strategies for articular cartilage and meniscus repair. Nat Rev Rheumatol. 2019;15:550–70.
4. Steadman JR, Rodkey WG, Rodrigo JJ. Microfracture: surgical technique and rehabilitation to treat chondral defects. Clin Orthop Relat Res. 2001;391:S362.

5. Hangody L, Vásárhelyi G, Hangody LR, Sükösd Z, Tibay G, Bartha L, et al. Autologous osteochondral grafting—technique and long-term results. Injury. 2008;39:32–9.
6. Hangody L, Kish G, Kárpáti Z, Udvarhelyi I, Szigeti I, Bély M. Mosaicplasty for the treatment of articular cartilage defects: application in clinical practice. Orthopedics. 1998;21:751–6.
7. Brittberg M, Lindahl A, Nilsson A, Ohlsson C, Isaksson O, Peterson L. Treatment of deep cartilage defects in the knee with autologous chondrocyte transplantation. N Engl J Med. 1994;331:889–95.
8. Brittberg M. Autologous chondrocyte implantation—Technique and long-term follow-up. Injury. 2008;39:40–9.
9. Nehrer S, Dorotka R, Domayer S, Stelzeneder D, Kotz R. Treatment of full-thickness chondral defects with hyalograft C in the knee: a prospective clinical case series with 2 to 7 years' follow-up. Am J Sports Med. 2009;37:81–7.
10. Anderson JA, Little D, Toth AP, Moorman CT, Tucker BS, Ciccotti MG, et al. Stem cell therapies for Knee cartilage repair: the current status of preclinical and clinical studies. Am J Sports Med. 2014;42:2253–61.
11. Wakitani S, Okabe T, Horibe S, Mitsuoka T, Saito M, Koyama T, et al. Safety of autologous bone marrow-derived mesenchymal stem cell transplantation for cartilage repair in 41 patients with 45 joints followed for up to 11 years and 5 months. J Tissue Eng Regen Med. 2011;5:146–50.
12. Yamasaki S, Mera H, Itokazu M, Hashimoto Y, Wakitani S. Cartilage repair with autologous bone marrow mesenchymal stem cell transplantation: review of preclinical and clinical studies. Cartilage. 2014;5:196–202.
13. Uccelli A, Moretta L, Pistoia V. Mesenchymal stem cells in health and disease. Nat Rev Immunol. 2008;8:726–36.
14. Gupta PK, Das AK, Chullikana A, Majumdar AS. Mesenchymal stem cells for cartilage repair in osteoarthritis. Stem Cell Res Ther. 2012;3:25.
15. Mackay AM, Beck SC, Murphy JM, Barry FP, Chichester CO, Pittenger MF. Chondrogenic differentiation of cultured human mesenchymal stem cells from marrow. Tissue Eng. 1998;4:415–28.
16. Koga H, Engebretsen L, Brinchmann JE, Muneta T, Sekiya I. Mesenchymal stem cell-based therapy for cartilage repair: a review. Knee Surg Sports Traumatol Arthrosc. 2009;17:1289–97.
17. Le H, Xu W, Zhuang X, Chang F, Wang Y, Ding J. Mesenchymal stem cells for cartilage regeneration. J Tissue Eng. 2020;11:2041731420943839.
18. Kangari P, Talaei-Khozani T, Razeghian-Jahromi I, Razmkhah M. Mesenchymal stem cells: amazing remedies for bone and cartilage defects. Stem Cell Res Ther. 2020;11:492.
19. Zha K, Li X, Yang Z, Tian G, Sun Z, Sui X, et al. Heterogeneity of mesenchymal stem cells in cartilage regeneration: from characterization to application. Npj Regen Med. 2021;6:1–15.
20. Jin HJ, Bae YK, Kim M, Kwon S-J, Jeon HB, Choi SJ, et al. Comparative analysis of human mesenchymal stem cells from bone marrow, adipose tissue, and umbilical cord blood as sources of cell therapy. Int J Mol Sci. 2013;14:17986–8001.
21. Dalecki D. Mechanical bioeffects of ultrasound. Annu Rev Biomed Eng. 2004;6:229–48.
22. Ling L, Feng X, Wei T, Wang Y, Wang Y, Zhang W, et al. Effects of low-intensity pulsed ultrasound (LIPUS)-pretreated human amnion-derived mesenchymal stem cell (hAD-MSC) transplantation on primary ovarian insufficiency in rats. Stem Cell Res Ther. 2017;8:283.
23. Zhou X, Castro NJ, Zhu W, Cui H, Aliabouzar M, Sarkar K, et al. Improved human bone marrow mesenchymal stem cell osteogenesis in 3D bioprinted tissue scaffolds with low intensity pulsed ultrasound stimulation. Sci Rep. 2016;6:32876.
24. Nakao J, Fujii Y, Kusuyama J, Bandow K, Kakimoto K, Ohnishi T, et al. Low-intensity pulsed ultrasound (LIPUS) inhibits LPS-induced inflammatory responses of osteoblasts through TLR4–MyD88 dissociation. Bone. 2014;58:17–25.
25. Zhou S, Schmelz A, Seufferlein T, Li Y, Zhao J, Bachem MG. Molecular mechanisms of low intensity pulsed ultrasound in human skin fibroblasts*. J Biol Chem. 2004;279:54463–9.
26. Ikai H, Tamura T, Watanabe T, Itou M, Sugaya A, Iwabuchi S, et al. Low-intensity pulsed ultrasound accelerates periodontal wound healing after flap surgery. J Periodontal Res. 2008;43:212–6.
27. Yuan Y, Wang Z, Wang X, Yan J, Liu M, Li X. Low-Intensity pulsed ultrasound stimulation induces coupling between ripple neural activity and hemodynamics in the mouse visual cortex. Cereb Cortex. 2019;29:3220–3.
28. Huang X, Niu L, Meng L, Lin Z, Zhou W, Liu X, et al. Transcranial low-intensity pulsed ultrasound stimulation induces neuronal autophagy. IEEE Trans Ultrason Ferroelectr Freq Control. 2021;68:46–53.
29. Malizos KN, Hantes ME, Protopappas V, Papachristos A. Low-intensity pulsed ultrasound for bone healing: an overview. Injury. 2006;37:S56-62.
30. Rutten S, van den Bekerom MPJ, Sierevelt IN, Nolte PA. Enhancement of bone-healing by low-intensity pulsed ultrasound: a systematic review. JBJS Rev. 2016;4: e6.
31. Min B-H, Choi BH, Park SR. Low intensity ultrasound as a supporter of cartilage regeneration and its engineering. Biotechnol Bioprocess Eng. 2007;12:22–31.
32. Tanaka E, Kuroda S, Horiuchi T, Tabata A, El-Bialy T. Low-intensity pulsed ultrasound in dentofacial tissue engineering. Ann Biomed Eng. 2015;43:871–86.
33. de Lucas B, Pérez LM, Bernal A, Gálvez BG. Ultrasound therapy: experiences and perspectives for regenerative medicine. Genes. 2020;11:1086.
34. El-Bialy T, Uludag H, Jomha N, Badylak SF. In vivo ultrasound-assisted tissue-engineered mandibular condyle: a pilot study in rabbits. Tissue Eng Part C Methods. 2010;16:1315–23.
35. Chauvel-Picard J, Korn P, Corbin S, Brosset S, Bera J-C, Gleizal A. Stimulation of oral mucosal regeneration by low intensity pulsed ultrasound: an in vivo study in a porcine model. J Prosthodont Res. 2021;65:46–51.
36. Budhiraja G, Sahu N, Subramanian A. Low-intensity ultrasound upregulates the expression of cyclin-D1 and promotes cellular proliferation in human mesenchymal stem cells. Biotechnol J. 2018;13:1700382.
37. Aliabouzar M, Lee S, Zhou X, Zhang GL, Sarkar K. Effects of scaffold microstructure and low intensity pulsed ultrasound on chondrogenic differentiation of human mesenchymal stem cells. Biotechnol Bioeng. 2018;115:495–506.
38. Xia P, Wang X, Wang Q, Wang X, Lin Q, Cheng K, et al. Low-intensity pulsed ultrasound promotes autophagy-mediated migration of mesenchymal stem cells and cartilage repair. Cell Transplant. 2021;30:0963689720986142.
39. Lai C-H, Chen S-C, Chiu L-H, Yang C-B, Tsai Y-H, Zuo CS, et al. Effects of low-intensity pulsed ultrasound, dexamethasone/TGF-β1 and/or BMP-2 on the transcriptional expression of genes in human mesenchymal stem cells: chondrogenic vs. osteogenic differentiation. Ultrasound Med Biol. 2010;36:1022–33.
40. Ebisawa K, Hata K, Okada K, Kimata K, Ueda M, Torii S, et al. Ultrasound enhances transforming growth factor β-mediated chondrocyte differentiation of human mesenchymal stem cells. Tissue Eng. 2004;10:921–9.
41. Takeuchi R, Ryo A, Komitsu N, Mikuni-Takagaki Y, Fukui A, Takagi Y, et al. Low-intensity pulsed ultrasound activates the phosphatidylinositol 3 kinase/Akt pathway and stimulates the growth of chondrocytes in three-dimensional cultures: a basic science study. Arthritis Res Ther. 2008;10:R77.
42. Mukai S, Ito H, Nakagawa Y, Akiyama H, Miyamoto M, Nakamura T. Transforming growth factor-β1 mediates the effects of low-intensity pulsed ultrasound in chondrocytes. Ultrasound Med Biol. 2005;31:1713–21.
43. Wang X, Lin Q, Zhang T, Wang X, Cheng K, Gao M, et al. Low-intensity pulsed ultrasound promotes chondrogenesis of mesenchymal stem cells via regulation of autophagy. Stem Cell Res Ther. 2019;10:41.
44. Louw TM, Budhiraja G, Viljoen HJ, Subramanian A. Mechanotransduction of ultrasound is frequency dependent below the cavitation threshold. Ultrasound Med Biol. 2013;39:1303–19.
45. Whitney NP, Lamb AC, Louw TM, Subramanian A. Integrin-mediated mechanotransduction pathway of low-intensity continuous ultrasound in human chondrocytes. Ultrasound Med Biol. 2012;38:1734–43.
46. Parvizi J, Parpura V, Greenleaf JF, Bolander ME. Calcium signaling is required for ultrasound-stimulated aggrecan synthesis by rat chondrocytes. J Orthop Res. 2002;20:51–7.
47. Chu CR, Szczodry M, Bruno S. Animal models for cartilage regeneration and repair. Tissue Eng Part B Rev. 2010;16:105–15.
48. Chung JY, Song M, Ha C-W, Kim J-A, Lee C-H, Park Y-B. Comparison of articular cartilage repair with different hydrogel-human umbilical cord blood-derived mesenchymal stem cell composites in a rat model. Stem Cell Res Ther. 2014;5:39.
49. Song B-W, Park J-H, Kim B, Lee S, Lim S, Kim SW, et al. A combinational therapy of articular cartilage defects: rapid and effective regeneration by using low-intensity focused ultrasound after adipose tissue-derived stem cell transplantation. Tissue Eng Regen Med. 2020;17:313–22.

50. Nishida T, Kubota S, Kojima S, Kuboki T, Nakao K, Kushibiki T, et al. Regeneration of defects in articular cartilage in Rat Knee joints by CCN2 (connective tissue growth factor). J Bone Miner Res. 2004;19:1308–19.
51. Azhari H. Basics of biomedical ultrasound for engineers. Hoboken: Wiley; 2010.
52. Huang X, Das R, Patel A, Duc NT. Physical stimulations for bone and cartilage regeneration. Regen Eng Transl Med. 2018;4:216–37.
53. Wojdasiewicz P, Poniatowski ŁA, Szukiewicz D. The role of inflammatory and anti-inflammatory cytokines in the pathogenesis of osteoarthritis. Mediators Inflamm. 2014;2014: 561459.
54. Hashimoto S, Nishiyama T, Hayashi S, Fujishiro T, Takebe K, Kanzaki N, et al. Role of p53 in human chondrocyte apoptosis in response to shear strain. Arthritis Rheum. 2009;60:2340–9.
55. Santangelo KS, Nuovo GJ, Bertone AL. In vivo reduction or blockade of interleukin-1β in primary osteoarthritis influences expression of mediators implicated in pathogenesis. Osteoarthr Cartil. 2012;20:1610–8.
56. Desai S, Jayasuriya CT. Implementation of endogenous and exogenous mesenchymal progenitor cells for skeletal tissue regeneration and repair. Bioengineering. 2020;7:86.
57. Embree MC, Chen M, Pylawka S, Kong D, Iwaoka GM, Kalajzic I, et al. Exploiting endogenous fibrocartilage stem cells to regenerate cartilage and repair joint injury. Nat Commun. 2016;7:13073.
58. Kobayashi S, Takebe T, Inui M, Iwai S, Kan H, Zheng Y-W, et al. Reconstruction of human elastic cartilage by a CD44+ CD90+ stem cell in the ear perichondrium. Proc Natl Acad Sci. 2011;108:14479–84.
59. Loyola-Sánchez A, Richardson J, Beattie KA, Otero-Fuentes C, Adachi JD, MacIntyre NJ. Effect of low-intensity pulsed ultrasound on the cartilage repair in people with mild to moderate knee osteoarthritis: a double-blinded, randomized, placebo-controlled pilot study. Arch Phys Med Rehabil. 2012;93:35–42.
60. Poolman RW, Agoritsas T, Siemieniuk RAC, Harris IA, Schipper IB, Mollon B, et al. Low intensity pulsed ultrasound (LIPUS) for bone healing: a clinical practice guideline. BMJ. 2017;356: j576.
61. Ren L, Yang Z, Song J, Wang Z, Deng F, Li W. Involvement of p38 MAPK pathway in low intensity pulsed ultrasound induced osteogenic differentiation of human periodontal ligament cells. Ultrasonics. 2013;53:686–90.
62. Inubushi T, Tanaka E, Rego EB, Ohtani J, Kawazoe A, Tanne K, et al. Ultrasound stimulation attenuates resorption of tooth root induced by experimental force application. Bone. 2013;53:497–506.
63. Chen K, Wang D, Du WT, Han Z-B, Ren H, Chi Y, et al. Human umbilical cord mesenchymal stem cells hUC-MSCs exert immunosuppressive activities through a PGE2-dependent mechanism. Clin Immunol. 2010;135:448–58.
64. Li T, Xia M, Gao Y, Chen Y, Xu Y. Human umbilical cord mesenchymal stem cells: an overview of their potential in cell-based therapy. Expert Opin Biol Ther. 2015;15:1293–306.
65. Xie Q, Liu R, Jiang J, Peng J, Yang C, Zhang W, et al. What is the impact of human umbilical cord mesenchymal stem cell transplantation on clinical treatment? Stem Cell Res Ther. 2020;11:519.
66. Xia P, Wang X, Qu Y, Lin Q, Cheng K, Gao M, et al. TGF-β1-induced chondrogenesis of bone marrow mesenchymal stem cells is promoted by low-intensity pulsed ultrasound through the integrin-mTOR signaling pathway. Stem Cell Res Ther. 2017;8:281.
67. Liu S, Zhou M, Li J, Hu B, Jiang D, Huang H, et al. LIPUS inhibited the expression of inflammatory factors and promoted the osteogenic differentiation capacity of hPDLCs by inhibiting the NF-κB signaling pathway. J Periodontal Res. 2020;55:125–40.
68. Xia P, Wang Q, Song J, Wang X, Wang X, Lin Q, et al. Low-intensity pulsed ultrasound enhances the efficacy of bone marrow-derived MSCs in osteoarthritis cartilage repair by regulating autophagy-mediated exosome release. Cartilage. 2022;13:19476035221093060.

## Publisher's Note